\documentstyle[12pt,aaspp4]{article}
\lefthead{Graham et al.}
\righthead{Brightest Cluster Galaxy Profile Shapes}
\begin{document}
\title{Brightest Cluster Galaxy Profile Shapes}
\author{Alister Graham,\altaffilmark{1} Tod R. Lauer,\altaffilmark{2,3} Matthew Colless,\altaffilmark{1} and Marc Postman\altaffilmark{4,5}}

\altaffiltext{1}{Mount Stromlo and Siding Spring Observatories, Australian National University, Private Bag, Weston Creek PO, ACT 2611, Australia.}
\altaffiltext{2}{Kitt Peak National Observatory (KPNO), National Optical Astronomy Observatories (NOAO), P.O. Box 26732, Tucson, AZ 85726.  NOAO is operated by the Association of Universities for Research in Astronomy (AURA), Inc., under cooperative agreement with the National Science foundation.}
\altaffiltext{3}{Visiting Astronomer, Cerro Tololo Inter-American Observatory (CTIO), NOAO.}
\altaffiltext{4}{Space Telescope Science Institute,\altaffilmark{6} 3700 San Martin Drive, Baltimore, MD 21218.}
\altaffiltext{5}{Visiting Astronomer, KPNO and CTIO.}
\altaffiltext{6}{Space Telescope Science Institute is operated by AURA, Inc., 
under contract to the National Aeronautics and Space Administration.}

\begin{abstract}
We model the surface brightness profiles of a sample of 119 Abell Brightest
Cluster Galaxies (BCG), finding a generalised deVaucouleurs R$^{1/n}$
law, where $n$ is a free parameter, to be appropriate.  Departures from the
$R^{1/4}$ law are shown to be a real feature of galaxy profiles, not due
to observational errors or coupling of $n$ with the other model parameters.
BCG typically have values of $n$ greater than 4.
The shape parameter $n$ is shown to correlate with effective half-light radius,
such that the larger BCG have larger values of $n$.   This 
continues a trend noticed amongst ordinary elliptical galaxies and dwarf 
ellipticals, such that the fainter galaxies have smaller values of $n$.



\end{abstract}


\section{Introduction}

The structure of the first-ranked members of Abell clusters,
or brightest cluster galaxies (BCG),
has long been of interest given that these galaxies can
be used as standard candles for exploring the large-scale structure
of the universe (\cite{HMS56}; \cite{AS72a}, \cite{AS72b};
\cite{GaO75}; Lauer \& Postman 1994). 
\cite{Hos80}, for example, developed the $L-\alpha$ relationship,
which relates BCG structure to their luminosities through
fixed physical apertures, thus reducing the cosmic scatter in the aperture
luminosities.
Understanding the $L-\alpha$ relationship has been of particular
interest recently, given its use by Lauer \& Postman (\cite{LP92}, 1994)
to probe the linearity of the Hubble flow and to detect large scale
bulk flows out to the 15,000 km/s scale.  The structure of BCG may also 
reflect galactic cannibalism (\cite{OaT75}; \cite{HaO78}; Hoessel 1980; 
\cite{SGH83}), where other cluster members may be captured through dynamical 
friction, changing both the luminosity and the shape of the dominant BCG.

While to first order BCG appear to be ordinary, if highly luminous,
elliptical galaxies, in many ways they are a special class of objects.
\cite{TaR77} showed, for example, that their
luminosity function was not consistent with them being drawn simply
as the brightest member of a standard \cite{Sch76} luminosity function.
\cite{Oem76} showed that as a class, BCG differed structurally from
other giant ellipticals of similar luminosity, having more extended
limiting radii at a given total luminosity than other giant
elliptical galaxies. 
\cite{Sbt86} conducted an extensive survey of BCG brightness
profiles, finding them to be shallower than those for more
ordinary elliptical galaxies, such that the BCG were always
more extended at a given surface brightness level than would
be suggested by simply scaling by total luminosity.  \cite{HOS87}
also concurred with this, showing BCG to follow a 
shallower (and tighter) relationship between effective radii, $R_e$, 
and effective surface brightness, $I_e$, than do ordinary elliptical 
galaxies (\cite{OaH91}; Graham 1996).
This distinction in form was additionally interesting as \cite{Sbt86}
showed it to arise in the relatively brighter mid-portions
of the BCG profile, regardless of whether or not the BCG had additional
envelopes at faint surface brightness making them cD galaxies as well.
A cD galaxy is a giant elliptical
that has a separate extended low surface-brightness envelope,
which is evident as an inflection in the brightness profile (Oemler 1976) ---
typically at $\mu_V\sim24$ or greater (\cite{KaD89}).

\cite{Sbt86} emphasized that the classic $r^{1/4}$ law, introduced by 
de Vaucouleurs (\cite{1948}, \cite{1953}) to describe elliptical galaxies at 
all luminosities, was often a poor match to the BCG profiles, in many
cases fitting the profiles only over a restricted range of
surface brightness.
Looking at the profiles presented by \cite{Sbt86}, it becomes
apparent that many of the BCG would be better fitted by power laws
rather than $r^{1/4}$ laws.
This is also found to be the case in Ledlow \& Owen's (1995) work with
galaxies in rich clusters.  This appears to have little to do with 
whether or not a BCG is also a cD galaxy.  
An additional caveat is that because a constant power law
will rise above an $r^{1/4}$ law at large radii, a cD envelope
may be erroneously detected as a separate component in $r^{1/4}$
plots, even though a single power law could describe the BCG completely.
In short, classification of a BCG as cD or not is problematic;
our present investigation, however, is not affected by this issue
as this apparently has little to do in any case with the
distinct over all structural properties of BCG from those of
giant ellipticals.

If some BCG are better described by power laws, and others by
$r^{1/4}$ laws, an inspection of the profiles presented
by \cite{Sbt86} shows that BCG fall along a continuum
between the two forms.
We are thus motivated to advance the \cite{Ser68} form, which
includes both laws as a better description and generalization
of BCG brightness profiles.
In the Sersic form, $I(r)\propto\exp(-(r/R_e)^{1/n}),$
where $n$ is a free parameter.
As $n\rightarrow1$ from the $n=4$ de Vaucouleurs form, the
Sersic form becomes increasingly exponential, while as $n\rightarrow\infty,$
the Sersic form becomes a pure power law.

\cite{YaC94} fitted the Sersic form to a sample
of dwarf elliptical (dE) galaxies, showing a correlation between galaxy
absolute magnitude and the shape parameter $n$, such that the fainter galaxies 
had the smaller values of $n$.
The same pattern in profile shape had also been noted by \cite{Dav88}. Working
with a sample of Fornax low surface brightness (LSB) diffuse dwarf elliptical 
and spheroidal galaxies, they showed a trend between $n$ and the logarithm
of the scale-radius $R_{e}$ such that the scale radius increased as $n$ 
increased.  Caon, Capaccioli, and D'Onofrio (1993) 
showed that this trend extended into the domain of ordinary 
elliptical galaxies.  It has also recently been shown to hold 
for the bulges of spiral galaxies by \cite{APB95} and \cite{CJB96}.

We are now in an excellent position to explore the systematics
of the BCG structural properties as \cite{PL95} present a complete and 
volume-limited sample of BCG out to 15,000 km/s.
The data was obtained with large-area CCDs and reduced with
photometric uniformity in mind, given the goal to use this data set
to detect the subtle effect of bulk flows on the $L-\alpha$
relationship residuals (Lauer \& Postman 1994).
Here we show that the correlation between $n$ and scale-radius 
(Caon, Capaccioli, and D'Onofrio 1993) extends to BCG, where $n$ is typically 
greater than 4, whereas for the dE it is seen to be less than 4.
The existence of this relation across such a large range of galaxy sizes
must be telling us something fundamental about the
structure and formation processes of elliptical galaxies in general.

In the next section we present some of the theory behind the $R^{1/n}$ model.
Section 3 presents the BCG surface brightness data and the parameterised model 
fits to it.
In Section 4 we discuss the results of the model fitting and compare this
in the context of work on other galaxies in Section 5.  We investigate
how $n$ relates to other measures of galaxy structure, both local, Section 6,
and global, Section 7.  Our conclusions are given in Section 8.

\section{The $R^{1/n}$ law} 
The $R^{1/n}$ law gives the observed galaxy intensity, I, as a function of 
radius such that
\begin{equation}
I(r)=I_{e}exp\left[ -b\left[\left( \frac{r}{r_{e}}\right) ^{1/n} -1\right]\right],
\end{equation}
where $I_{e}$ is the intensity at the radius $r_{e}$.  The constant b is
chosen so that $r_{e}$ becomes the radius enclosing half of the
total light from the galaxy.
This generalised deVaucouleurs law was introduced by \cite{Ser68} and 
further developed by \cite{Cio91}, the
deVaucouleurs law has n=4 and b=7.67. Note that the constant b in this 
formula is a function of $n$.  The generalised expression can be written as 
\begin{equation}
\mu (r)=\mu _{0}+\frac{2.5b_{n}}{ln(10)}\left(r/r_{e}\right)^{1/n}, \label{r2n}
\end{equation}
with $r_{e}$ the scale radius, $\mu _{o}$ is the central surface brightness
 and $b_{n}$ the function of $n$ given below. Again we select $b_{n}$ 
such that $r_{e}$ is the radius enclosing half of the total light for the 
$R^{1/n}$ model.  For n=4, the well known deVaucouleurs $R^{1/4}$ formula, 
$\mu =\mu _{0}+8.33(r/r_{e})^{1/4}$ is recovered.  
The luminosity interior to the radius r is given by
\begin{equation}
L(r)=I_{e}r_{e}^{2}2\pi n(1-\epsilon )\frac{e^{b_{n}}}{(b_{n})^{2n}}\gamma (2n,x), \label{lumin}
\end{equation}
for galaxy ellipticity $\epsilon $, and where $\gamma $(2n,x) is the 
incomplete gamma function with $x=b_{n}(r/r_{e})^{1/n}$, defined by
\begin{equation}
\gamma (2n,x)=\int ^{x}_{0} e^{-t}t^{2n-1}dt.
\end{equation}
Thus, the value of $b_{n}$ is such that $\Gamma (2n)=\gamma (2n,b_{n})$, where 
$\Gamma $ is the gamma function.  As given by \cite{Cap89}, this can be well 
approximated by $b_{n}=1.9992n-0.3271$.

The differences between an $R^{1/n}$ model and an $R^{1/4}$ model are best
described graphically.  In Figure~\ref{udiff}, we plot these differences for
n ranging from 1 to 10.  The models have been 
constructed to have the same total luminosity, in units of $I_{e}$.
The abscissa is in units of the half-light radius of the $R^{1/4}$ model.
As $n$ climbs greater than 4, the curvature in the profile is
steadily removed until it approximates a straight line.  
As $n$ exceeds 10, the differences in the profiles become less marked.
For these high values of $n$, the the portion of the model profile that is 
matched by observations is well represented by a power law. 
It is seen that for the range in radius that can be matched by observations,
values of $n<4$ lead to a hump in the profile indicating that the $R^{1/4}$ 
profile is too faint for the central parts and too bright to match the inner
and outer portions of the profile.  The situation is reversed for $n>4$,
such that the brighter galaxies have more light at larger radii than 
the $R^{1/4}$ law allows.

\section{Observational surface brightness profiles and fitted models}

The galaxies studied here are the BCG of the volume-limited sample
of the 119 Abell clusters known to be within 15,000 km/s, which were selected
by \cite{LP94} to define a large-scale inertial reference frame.
Full details of the galaxy selection, observations, and data reduction
are given in \cite{PL95}, but we summarize them briefly here.
BCG candidates were selected from sky-survey images, and imaged
in the Kron-Cousins R band under photometric conditions with large-area
CCD cameras at KPNO and CTIO.
The distance indicator used by \cite{LP94} is based on the
integrated luminosity of the BCG within the central 10h$^{-1}$ Kpc
metric aperture, thus final selection of a given galaxy as ``brightest''
among rival candidates is by metric rather than total luminosity.
On this note, \cite{PL95} emphasize that a number of
their BCG identifications differ from those made earlier by \cite{Hos80}.

\cite{PL95} measured profiles of the BCG using the
multi-isophote fitting algorithm of \cite{Tod86}.
Many of the BCG are part of multiple-galaxy systems;
the multi-isophote algorithm solves for the brightness distributions
of all overlapping galaxies simultaneously.
Compact galaxies, stars, CCD defects, and so on, can also be excluded
from the isophote fitting algorithm.
Final accuracy of the profiles is limited by photometric calibration
at small radii, and sky subtraction errors at large radii.
\cite{PL95} used repeat observations of the same
galaxies to show that the basic random error in the BCG aperture
magnitudes is only 0.014 mag; the surface brightness
values for isophotes within the metric radius
will be accurate to a similar level.
Errors in the photometry due to sky subtraction errors will begin
to dominate outside the metric radius, but should still be at the
few percent level at the limiting isophote brightness adopted
here of $\mu_R=23.5.$
\cite{PL95} did sky subtraction using a mode-estimator
in the image corners.  Some BCG were still contributing
significant amounts of light at the edges of even the large fields of the CCDs
used, in which cases additional observations were obtained
offset from the primary galaxy image to measure the sky at larger
angular distances from the BCG.
Lastly, we note that the galaxy profiles are presented as observed.
No $K$ or extinction corrections have been applied (although \cite{PL95}
did apply them to derive BCG metric absolute luminosities).

The model profiles have been fitted to the data outside of the central 3 arc 
seconds, due to the possible influence of core structure that is separate to 
the outer galaxy profile.  We also have not used the profile data fainter than 
23.5 magnitudes to be sure we are not affected by sky subtraction errors.  In 
addition, this level of truncation in the profile ensures that our results are 
not a product of the extended haloes or envelopes that cD galaxies are
known to have.  For
many of the BCG we had multiple images and could directly compare different
profiles of the same galaxy for agreement.  The model parameters obtained
for galaxies with multiple images are found to be in agreement with each
other within the errors.

Equation~\ref{r2n} has been fitted to the semi-major axis surface brightness 
profiles of the BCG, using a simple error weighting scheme based on the S/N 
of each data point, via standard non-linear least-squares to solve for the 
three unknowns.  We give here the superscript $n$ to the value 
of $\mu _{0}$ and $r_{e}$ derived from the $R^{1/n}$ formula, to prevent 
confusion with the values derived from the $R^{1/4}$ formula.
$\Delta \chi ^{2}$ ellipsoids were computed around the best fitting 
parameters, $\mu _{0}^{n}, r_{e}^{n}$, and $n$.  This was done by moving 
through a fine 3D grid of values and computing the value of $\chi ^{2}$ at 
each point.  $r_{e}$ was converted to $\log r_{e}$ and the ellipsoids 
then projected onto the relevant 2D plane.

In addition, we fitted for the standard $R^{1/4}$ formula, where the value of n
is fixed at 4 and one solves for $\mu _{0}$ and $r_{e}$.
We also explored the use of a power law to describe the light profiles.
The method of least-squares was used to fit the profiles to
\begin{equation}
\mu (r\arcsec )=A+B\log (r\arcsec ),
\end{equation}
where A is referred to as the intercept and is approximately the central 
surface brightness, being the value at $r=1\arcsec $.  B is the slope of this 
power law.  We divide B by -2.5 so that it reflects the slope of the 
log(Intensity)-log(radius) profile.  Future references will refer to this
modified value of B as being the slope.
For the very flattened profiles, $n>15$, the profiles lacked any significant
curvature and it was appropriate to fit a power law of the above form 
to the data.

\section{BCG profiles}

The various profiles, $R^{1/4}, R^{1/n}$ and a simple power law, have been
fitted to the BCG surface brightness profiles and are displayed in 
Appendix A.  The residuals of the data about these best fitting models is 
shown clearly in Figure~\ref{ProRes} for a handful of galaxies, as are the 
measurement uncertainties, based on S/N measurements, 
associated with the data.  In general, one finds that the $R^{1/4}$ law 
has too much curvature to match the data, resulting in a negative bowl shaped
residual profile.  The opposite is found for the fitting of a power law,
where one generally finds the profile data has some level of curvature that 
cannot be accounted for with a simple power law, resulting in a positive hump
in the residual profile.  The $R^{1/n}$ model with its free shape parameter
can account for the differing levels of curvature and thus provides the
best fit to the data, ironing out the large scale departures seen in the
above two models.

Shown in Figure~\ref{nvsRe} are the best values of $n$ plotted against 
$\log r_{e}^{n}$ for the BCG.  Also shown are the projected 
$\Delta \chi ^{2}=9.21$ ellipses about these points, 
corresponding to a $99\%$ confidence region (\cite{Pre86}).
It is pointed out that the profiles have additional wiggles in them that 
are not accommodated for by the $R^{1/n}$ law and these exist at a level
greater than the observational errors.  As a result, the reduced $\chi ^{2}$
values for each profiles optimal fit is larger than 1.  Given that we
have not under-estimated our errors, as shown by the good agreement of repeat
measurements, this would imply that the model being fitted is inadequate. 
It is true that there are features/wiggles in the profiles that are not
explained by the $R^{1/n}$ law, nor are they explained by the $R^{1/4}$
law.  However, the bulk shape of the profile is described 
better by an $R^{1/n}$ law than an $R^{1/4}$ law, as indicated in 
Table 1 which shows the reduced $\chi ^{2}$ values for the models 
fitted to the data.  This is illustrated in Figure~\ref{figX},
which has the reduced $\chi ^{2}$ values of each model's optimal fit 
plotted against the value of $n$ from the $R^{1/n}$ profile fit. One can see
some general trends present.  As one would expect, the $R^{1/4}$ profile 
fits are best
when the $R^{1/n}$ fits have $n$=4.  The power law fits are better for larger
values of $n$, which is to be expected as larger values of $n$ mean a profile
with less curvature.  The $R^{1/n}$ fits are better than those of the
other two models and the quality of the fit appears to be independent of $n$.
As an alternative measure of the errors, we set the reduced 
$\chi ^{2}=1$ for the optimal fit and computed the 1$\sigma $ ellipses 
normalised to this level.  These are also shown in Figure~\ref{nvsRe}.

Seeing effects are not responsible for the observed correlation of $n$
with radius.  We explored excluding different inner portions of the light
profile.  \cite{Sag93} showed that the effects of seeing on the 
photometric properties of elliptical galaxies can extend as far as 
5 seeing discs.  Not using the inner $7.5\arcsec $, the same general trend 
between $n$ and $\log r_{e}$ is still obtained.

Not surprisingly, the $R^{1/n}$ model is the best performer, as it can
represent the $R^{1/4}$ profile when n=4 and it can approximate a power
law for large values of $n$, and fit for profiles of intermediate type. 
This can be attributed to having 1 more free parameter than the $R^{1/4}$
or power law.  What is important, is that such a variety in profiles are 
real, as indicated by the
error ellipses in Figure~\ref{nvsRe}.  To explore this further, we simulated 
a pure $R^{1/4}$ law profile out to 1$R_{e}$ and added to it random gaussian 
noise, with standard deviations varying as a function of radius and being
derived from the mean S/N errors of the 119 BCG.  We ran a Monte
Carlo simulation, solving for $r_{e}^{n}, \mu _{0}^{n}$, and $n$ each time
with a new set of random noise to see if we could explain the range of 
values in $r_{e}^{n}$ and $n$ observed with the real data.  
Figure~\ref{monte} shows the entire cloud of solutions, not just the 1 or 2 
$\sigma $ ellipses.  Clearly observational noise cannot explain the 
claimed trend of $n$ with $r_{e}^{n}$, suggesting that it is physically real 
and galaxies do indeed exhibit such 
a range in profile shapes.  The possibility, mentioned by Kormendy
(\cite{1980,1982}), that such a correlation may not be physically significant
but rather due to coupling of the parameters in the fitting formula does not
fully explain the observed variation in $n$, or the correlation of $r_{e}$
with $n$.
This analysis does however exclude the influence of sky subtraction errors,
which are estimated to be at a level of 0.3$\%$, and should not be 
significant.  In addition, the S/N weighting scheme, employed in the model
fitting, will down play possible contributions from sky subtraction errors.

In fitting the $R^{1/n}$ formula,
it was not possible to obtain meaningful results for all profiles.
The flattened profiles had large values of $n$, and implausibly large 
half-light radii.  As the shape parameter increased, it lost its sensitivity
and for values of $n$ greater than about 15, the profiles resembled straight 
lines. That is, the profiles were better described by a power law.
For about 40$\%$ of the data it was not appropriate to fit an $R^{1/n}$ law,
but rather a power law.  This can be compared with the work of \cite{LaO95}
who found for a control sample of 50 non-radio selected 
galaxies in rich clusters, 39$\%$ preferred a power law fit.
The slope and intercept of the best fitting power law for each BCG is plotted
in Figure~\ref{Power}.  Not surprisingly, there is a slight trend for galaxies 
with a steeper profile to have a brighter central flux, represented by the 
$\mu $ intercept at r=1$\arcsec$.  As the power law is appropriate for BCG 
with $n>15$ these galaxies have been marked with a star so as to distinguish 
them from the other galaxies that are not well represented by a power law.
Both types occupy similar regions, with the BCG well fitted by a power law not
occupying any special domain of this diagram.
Of course the real galaxy profile must turn over and be truncated at some 
radius, otherwise it would be of infinite size and luminosity.  But for the 
range in radius explored, some 40$\%$ of the BCG had
surface brightness profiles of a power law nature.  Ordinary elliptical 
galaxies and certainly dE do not show this feature over the same range in 
radius.  It is pointed out that for large values of $n$, the half-light
radius obtained from the $R^{1/n}$ model is not really physical, but is
however an expression of the slope of the galaxy profile measured by the 
data.

We find no correlation between $n$ and ellipticity.  Following \cite{CCDet},
we have plotted $n$ against the maximum ellipticity of the galaxy in
Figure~\ref{ellip}.  We also used the ellipticity at the semi-major axis 
radius of 10 {\it h}$^{-1}$ Kpc and again found no evidence for a correlation.
\cite{CCDet} found a range of ellipticities for $n<4$ but for $n>4$ the 
ellipticity was typically less than 0.3.  A possible reason for this 
different result could be that we used BCG whilst \cite{CCDet} 
used E/S0 galaxies, indicating a difference between the galaxy classes. 
Although this seems hard to understand, we note that \cite{CCDet} only had 
a dozen galaxies with $n>4$ and only two with $n>10$. They also found that 
this trend disappeared when they used the value of $n$ constructed from the 
semi-minor axis profile.  

Figure~\ref{enmag} shows $n$ plotted against the metric magnitude, being that
enclosed by a circular aperture of 12.5 Kpc (we used a value of 
$H_{0}=80$ km s$^{-1}$ Mpc$^{-1}$, and use the CMB reference frame here).  
A linear correlation coefficient of
-0.17, at a significance level of 84\%, suggest that there is little
correlation between these two values for our BCG sample.  This is not
surprising given the small scatter in metric magnitudes for the BCG, being
only 0.33 magnitudes.  That the BCG metric magnitude is not significantly 
influenced by the global galaxy profile shape, represented by $n$, eliminates
the chances of another hidden variable giving scatter to this magnitude
which has been used as a distance indicator by \cite{LP94} and \cite{Col95}
in their studies of peculiar velocity flows.

\section{$R^{1/n}$ Universality}

What influences the profile shape of galaxies?
Is it because of environment, changes arising due to dynamical evolution, 
differing formation history, or individual galaxy peculiarities such as 
dust clouds or rings, lenses, shells, ripples, etc.?  
The trend of $n$ with galaxy size argues against galaxy peculiarities 
being responsible, as this would require such features to be correlated
with galaxy size.  The lack of a disk for the BCG eliminates the 
option that a disk is the responsible feature in determining the galaxy shape.
However, the small wiggles in the profiles that are not accommodated by the
$R^{1/n}$ law are at a significance level greater than accounted for by
the observational uncertainty and may be signs of such features.
A close analysis of the two dimensional residual maps (images with 
the best fitting $R^{1/n}$ law subtracted) could reveal such features.

To investigate possible environmental effects on the BCG profile shape, we 
checked for a correlation between $n$ and Richness Class, RC, (\cite{Abe58}) 
and $n$ and Bautz-Morgan type, BM, (\cite{BaM70}).  We only used those galaxies
which had values of $n<15$, being some 60\% of our sample.
No obvious trend was found for either case, as can be seen in 
Figure~\ref{FigRC} and Figure~\ref{FigBM} respectively.  We point out that the
profile shape has been determined from the inner surface photometry 
($<23.5$ mag) and is thus free from outer galaxy distortions such as 
envelopes, which may be influenced by the environment.
\cite{Ein93} have shown $n$ not to correlate with galaxian density for their
sample of E/S0 galaxies; 
therefore further restricting the possibility of environment being responsible
for the shape parameter $n$.  

BCG typically like values of $n$ greater than 4.  This is not only interesting
in itself, but becomes more so when we note that studies of dE galaxies 
show them to have values of $n$ typically less than 4.  \cite{CCDet} show that 
ordinary elliptical and S0 galaxies show a range in $n$ above and below 4.
Continuing figure 5 of \cite{CCDet} we add our data to produce 
Figure~\ref{master}, showing the continued relationship between galaxy 
structure and size for dwarf galaxies, ordinary E/S0 galaxies and BCG. 
\cite{CCDet} fitted the $R^{1/n}$ law to the B-band photometry of 33 E/S0 
galaxies.  Whilst we used the Kron-Cousins R-band to image our sample
of BCG, changes in profile shape due to color gradients are 
expected to be lost in the scatter of Figure~\ref{master}.  We note that
the values of $R_{e}$ from \cite{CCDet} are not those from their model
but are derived directly from the galaxy light profile.
The dwarf galaxies are a sample of 187 Fornax cluster dwarfs
from \cite{Dav88}.  They fitted a variant form of the $R^{1/n}$ law to the
B-band images, such that $I(r)=I_{0}exp[-(r/A)^N]$.  By setting 
N=1/n, one has $R_{e}\arcsec =A\arcsec (2/N-0.327)^{1/N}$.  We used 
a value of $H_{0}=80$ km s$^{-1}$ Mpc$^{-1}$ throughout in determining
the value of $R_{e}$ Kpc.

We thus have a parameter that traces structural differences
amongst galaxies having a large range in size.  The fact that the $R^{1/n}$
profile is applicable over such a range (6 orders of magnitude in the mass)
suggests that there are similar formation mechanisms present for all these
galaxies.  Computer modelling of the physics of violent relaxation has shown 
that the shape parameter $n$ is dependent on the central potential of the 
models (\cite{HaM95}).  Galaxies with larger potentials
result in galaxies with larger values of $n$, and those with smaller
potentials have values of $n$ less than 4, in agreement with what the 
observations reveal.  This suggests that the bulk shape to an elliptical
galaxies luminosity profile is not due to its individual genetic peculiarities 
or its 
environment but rather something intrinsic to each galaxy, being its mass.  
This has previously been suggested by \cite{YaC94} and \cite{APB95}.
We hope to investigate this further when we have velocity dispersion 
data for the BCG.  This information can be used to estimate the mass
of each galaxy (\cite{VBS95,Mic80}), which we can then check for a 
correlation with $n$.

\section{Local profile structure $\alpha $}

The shape parameter, $n$, describes the overall global shape of the galaxy 
light profile.  There is another measure of a galaxies light profile 
curvature, referred to as the structure parameter by \cite{Hos80} and given
by $\alpha =d\ln L/d\ln 
r|_{r_{s}}$.  It is a measure of the galaxy profile at a given sampling
radius, $r_{s}$.  \cite{LP94} and \cite{Col95} have used 
$\alpha $ because of its correlation with metric magnitude and hence ability 
to create improved standard candles for distance measurements.  In each
case, $\alpha $ was measured directly from the light profiles at a sampling
radius of 10h$^{-1}$ Kpc.  It is, however, possible for one to compute 
$\alpha $ from the fitted model profile \cite{Gra96}.  Whether this is
preferred or not, one gains insight into how $\alpha $ varies for different 
values of $n$.  Working from equation~\ref{lumin} and assuming zero 
ellipticity for simplicity, it can be shown that 
\begin{equation}
\alpha =\frac{e^{-x}x^{2n}}{n\gamma (2n,x)}.
\end{equation}

A set of $R^{1/n}$ profiles were constructed and $\alpha $ computed as
a function of radius for each.  The results of which are shown in 
Figure~\ref{alp}.  For a fixed sampling radius, $r_{s}$, $\alpha $ is
seen to increase as the effective half-light radius increases.  Working 
against this is the fact that as galaxies get bigger, ie. larger $R_{e}$,
they tend to be described by a profile with a larger value of $n$, as seen 
in Figure~\ref{nvsRe}.  Now profiles with a larger values of $n$, are seen to 
have smaller values of $\alpha $ for the same ratio of sampling radius to 
effective half-light radius.  The degree to which $\alpha $ changes for
a given change in sampling radius or change of $n$ is dependent on the part
of the profile one is looking at.

We have plotted the values of $n$ against $\alpha $ for our sample of BCG 
that have values of $n<15$ in Figure~\ref{enalp}.
Values of $\alpha $ have been taken from \cite{LP94} rather than computed
from the fitted profile model. 
It seems likely that the reason no trend is evident between 
these two parameters is because of the competing situation described above.

\section{Another shape parameter}

An alternative approach to quantifying the systematic deviations from an
$R^{1/4}$ law was taken by \cite{Bur93}.  The observational data in the 
range $0.6<x<1.1$ was fitted by the form $\mu (x)=\mu _{0}+bx$, 
where $x=(r/X_{e})^{1/4}$ and $X_{e}$ is the effective half-light radius of 
the model.  In general, $\mu _{0}$ and b are dependent on the fitted range 
in x, which is in turn 
dependent on the value of $X_{e}$.  As we don't have a prior knowledge of the 
value of $X_{e}$, this dependence means that our expression must be 
solved via an iterative process until convergence, such that 
$X_{e}^{new}=X_{e}^{old}(8.3268/b)^{4}$, with b the best fitting slope 
using $X_{e}^{old}$.

Burkert observed that many galaxy profiles had a characteristic dip/hump about 
the best fitting $R^{1/4}$ law.  When the dip $(\mu _{galaxy}-R^{1/4})$ was a 
minimum,
it was found to turn over near x=0.8, and when it was a maximum it would be 
near x=0.9.  See \cite{Bur93} for a fuller description and profiles. The 
above mentioned range in x was then cut into two parts, with the division at 
$x_{cut}=0.8$ if the dip was a minimum, or at $x_{cut}=0.9$ if it was a 
maximum.
\begin{equation}
\delta b=\left(\frac{\partial u}{\partial x}\mid _{x\in [x_{cut};1.1]} - \frac{\partial u}{\partial x}\mid _{x\in [0.6;x_{cut}]}\right)/8.3268
\end{equation}
was shown to correlate with both absolute magnitude and the mean deviation of
the galaxy data about the best fitting $R^{1/4}$ law.  Burkert showed that the
brighter galaxies had the more negative values of $\delta b$, and the fainter
galaxies had the more positive values.

This parameter reflects the curvature present in the galaxy profiles that
the $R^{1/4}$ law can't accommodate.  However, it's extraction is somewhat 
fiddly and requires the fitting function be applied several times, whereas 
the $R^{1/n}$ profile is fitted once.  A series of $R^{1/n}$ profiles were 
created, and the method of Burkert applied to obtain the parameter $\delta b$
for each.  The relation between $n$ and $\delta b$ is shown in 
Figure~\ref{dbvsn}.
The sense and amplitude of $n$ and $\delta b$ are in agreement, such that the
bigger brighter galaxies with $n>4$ relate to $\delta b<0$ and the galaxies
with $n<4$ have $\delta b>0$.

\section{Conclusions}

The nature of the $R^{1/n}$ law is such that as $n$ increases the galaxy 
surface brightness profile flattens, and for $n>15$ it is very well 
represented by a power law.  
BCG typically have values of $n$ greater than 4 - that is their light profiles
are less curved than the classic $R^{1/4}$ law; 40$\%$ of our sample is
well described by a power law.
The range in profile shapes is real and not due to noise in the galaxy 
profiles or due to a coupling of the three parameters in the model.
There is a trend between $n$ and the half-light radius, such that the
larger galaxies have larger values of $n$.
This trend appears to be a continuation of that noticed for dE galaxies, LSB 
galaxies, through to normal E and S0 galaxies and on to BCG, suggesting 
some common physical processes must be at play in the formation of all
of these galaxies.  This global shape parameter, $n$, is shown to be
independent of Richness Class and Bautz-Morgan type, suggesting that 
the galaxy environment (in so far as RC and BM type represent this) 
is not responsible for the shaping of the bulk
distribution of stars in the galaxy.  We note that our analysis excludes 
the outer envelopes of the cD galaxies.


Whilst $n$ is a global measure of the galaxies light profile,
$\alpha $ is a measure the galaxies structure at a point.  $\alpha $ is shown 
to be related to $n$ and $R_{e}$ in an opposing manner.   As one moves to 
larger galaxies, $n$ increases causing $\alpha $ to decrease but at the same
time $R_{e}$ increases causing $\alpha $ to increase.  The dominating 
factor depends on which part of the profile one samples the value of 
$\alpha $ at.

\acknowledgments

Roberto Saglia is thanked for his comments and suggestions on this paper.
Joe Morris is thanked for help given during the preparation of this work.

\appendix
\section{Brightest Cluster Galaxy Surface Brightness Profiles}

The surface brightness profile data for all 119 BCG, together with the best 
fitting $R^{1/4}$ (dotted line), $R^{1/n}$ (solid line), and power law 
(dashed line) are displayed.  The models were fitted to the error weighted 
data brighter than 23.5 magnitudes (Kron-Cousins R band) and outside 3 
arcseconds (filled circles). Data outside this range being given by an open
circle.  The S/N errors are shown, but usually are contained within the
data point symbol. The surface brightness level at 22.0 magnitudes is indicated
for each profile by a small arrow on the right hand margin.  
No $R^{1/n}$ profile is shown when $n>15$.

\clearpage

\begin{figure}
\figcaption{a) Surface brightness profiles of an $R^{1/n}$ law, for integer values of $n$ ranging from 1 to 10.  b) The difference between the surface brightness profiles of an $R^{1/n}$ and an $R^{1/4}$ law, for $n$ ranging from 1 to 10 in steps of 1.  All profiles give the same total magnitude, in units of $I_{e}^{n}$.\label{udiff}} 
\figcaption{We show the surface brightness profiles for a sample of BCG, together with the best fitting $R^{1/4},$ (dotted line) $R^{1/n}$ (solid line) and power law (dashed line).  The residuals of the profiles about these models are displayed in a magnified portion below the profiles themselves, being the circles, stars and triangles for the respective models.\label{ProRes}}
\figcaption{upper panel) Plot of $n$ versus $\log r_{e}^{n}$ (Kpc) for our sample of galaxies.  (40$\%$ of the BCG had values of $n>15$). middle panel) $99\%$ confidence regions. lower panel) $1\sigma $ ellipses after the reduced $\chi ^{2}$ has been normalised to 1.\label{nvsRe}} 
\figcaption{Reduced $\chi ^{2}$ values from the optimal fit of the $R^{1/4}$, power law, and $R^{1/n}$ models to the BCG profiles are shown here against the value of $n$ from the optimal $R^{1/n}$ fit.\label{figX}}
\figcaption{Cloud of solutions from a Monte Carlo investigation of the contribution to the $n-\log r_{e}^{n}$ relation from observational errors.\label{monte}} 
\figcaption{Plot of the best fitting power law slope and intercept at $r=1\arcsec$ for the BCG.   Those galaxies fitted with a value of $n>15$ from the $R^{1/n}$ model have been marked with a star, those fitted with a value of $n<15$ are marked with a square.  The $1\sigma $ error bars being smaller than the graph markers.\label{Power}}
\figcaption{Plot of $n$ against the maximum ellipticity of the BCG.\label{ellip}}
\figcaption{Plot of $n$ versus the metric magnitude contained with 10h$^{-1}$ Kpc.\label{enmag}}
\figcaption{The galaxy shape parameter, $n$, is plotted against the galaxy cluster Richness Class, RC.  Also shown are the mean values for each class and the associated 1$\sigma $ confidence interval.\label{FigRC}}
\figcaption{The galaxy shape parameter, $n$, is plotted against the Bautz-Morgan, BM, type for the galaxy cluster.   Also shown are the mean values for each BM type and the associated 1$\sigma $ confidence interval.\label{FigBM}}
\figcaption{The galaxy shape parameter, $n$, is plotted against the logarithm of the effective half-light radius, $R_{e}$ Kpc, for a sample of dwarf galaxies (Davies et al.\ 1988), triangles, ordinary E/S0 galaxies (Caon et al.\ 1993), filled circles, and our BCG data, open circles.\label{master}}
\figcaption{The structure parameter, $\alpha $, is plotted as a function of radius for different values of the shape parameter, $n$, from the $R^{1/n}$ model.  $r_{s}/r_{e}$ is the ratio of the sampling radius, at which $\alpha $ is computed, to the effective half-light radius of each respective model.\label{alp}}
\figcaption{Plot of $n$ versus $\alpha $, where $\alpha $ is taken from Lauer \& Postman (1994).\label{enalp}}
\figcaption{A series of $R^{1/n}$ profiles were constructed, from which we computed $\delta b$ (Burkert 1993).  The relation between these two measures of the galaxy light profile curvature are shown.\label{dbvsn}} 
\end{figure}

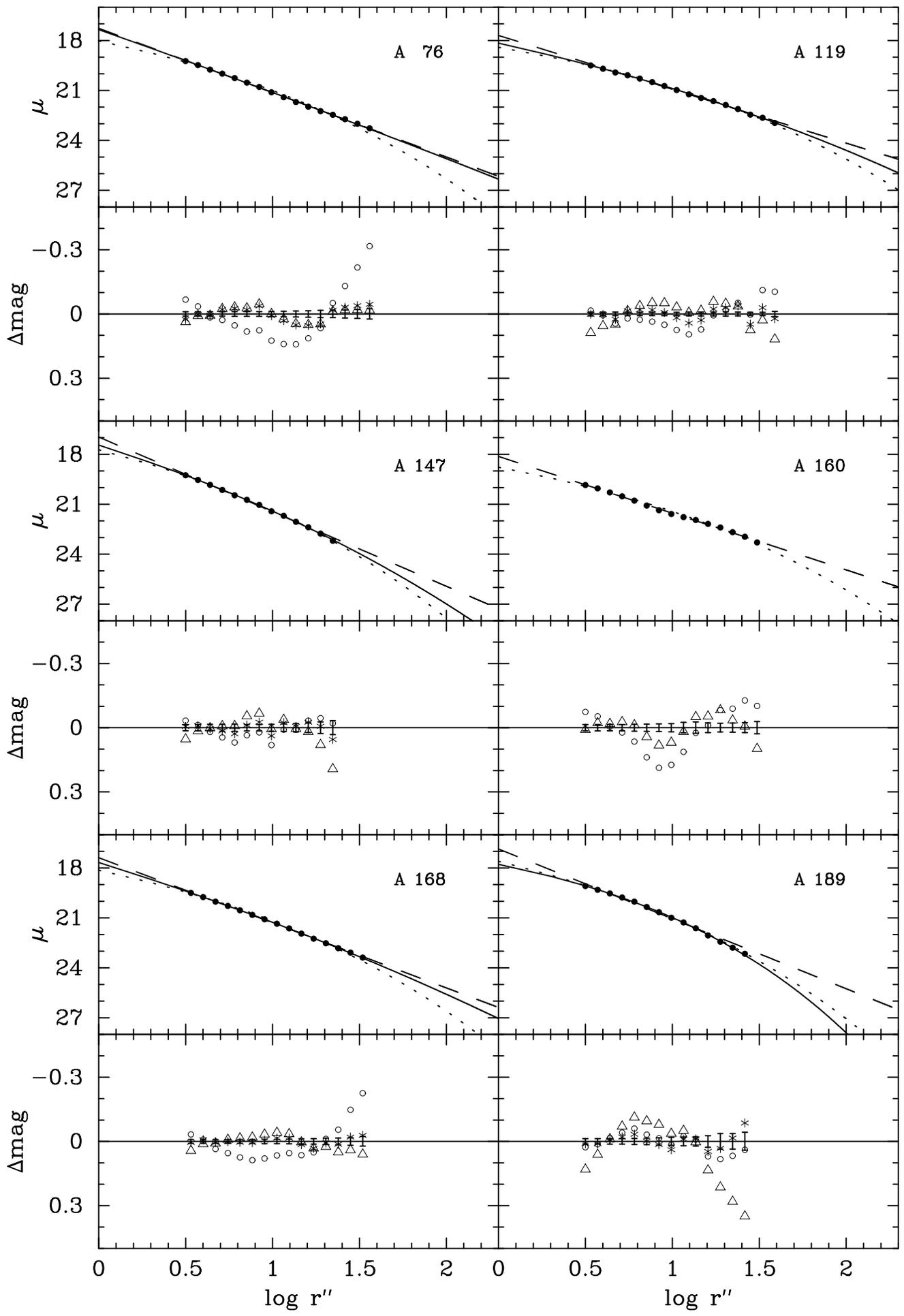
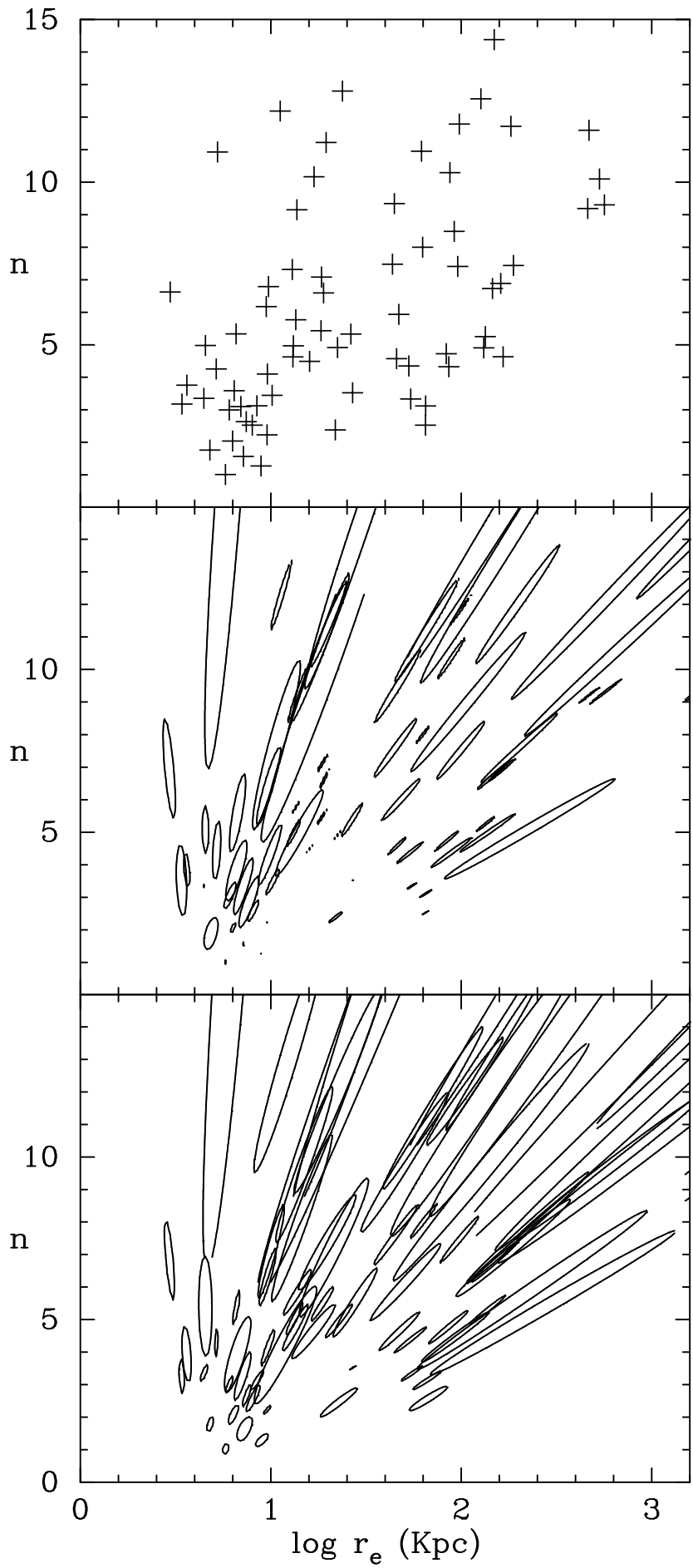
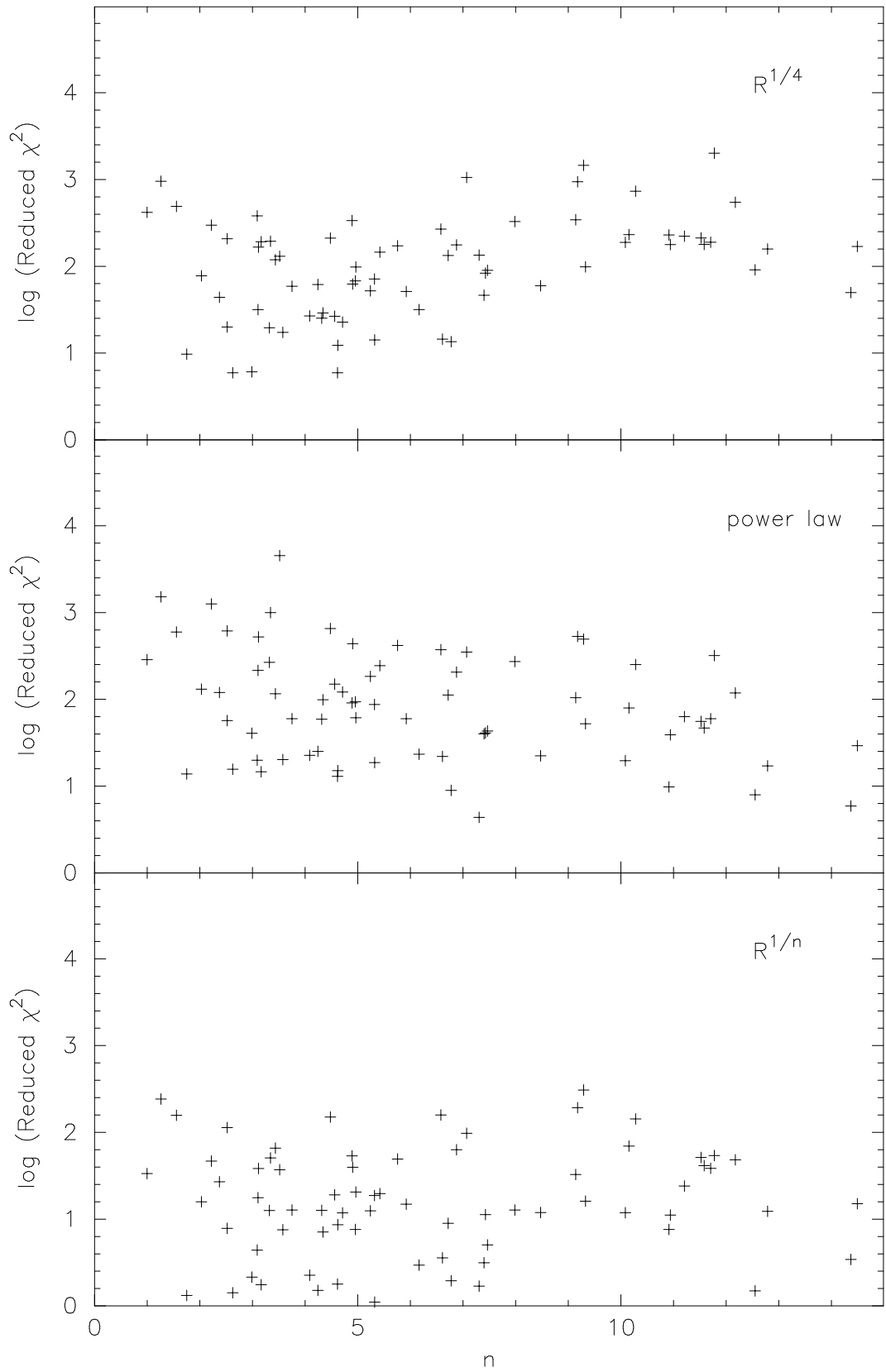
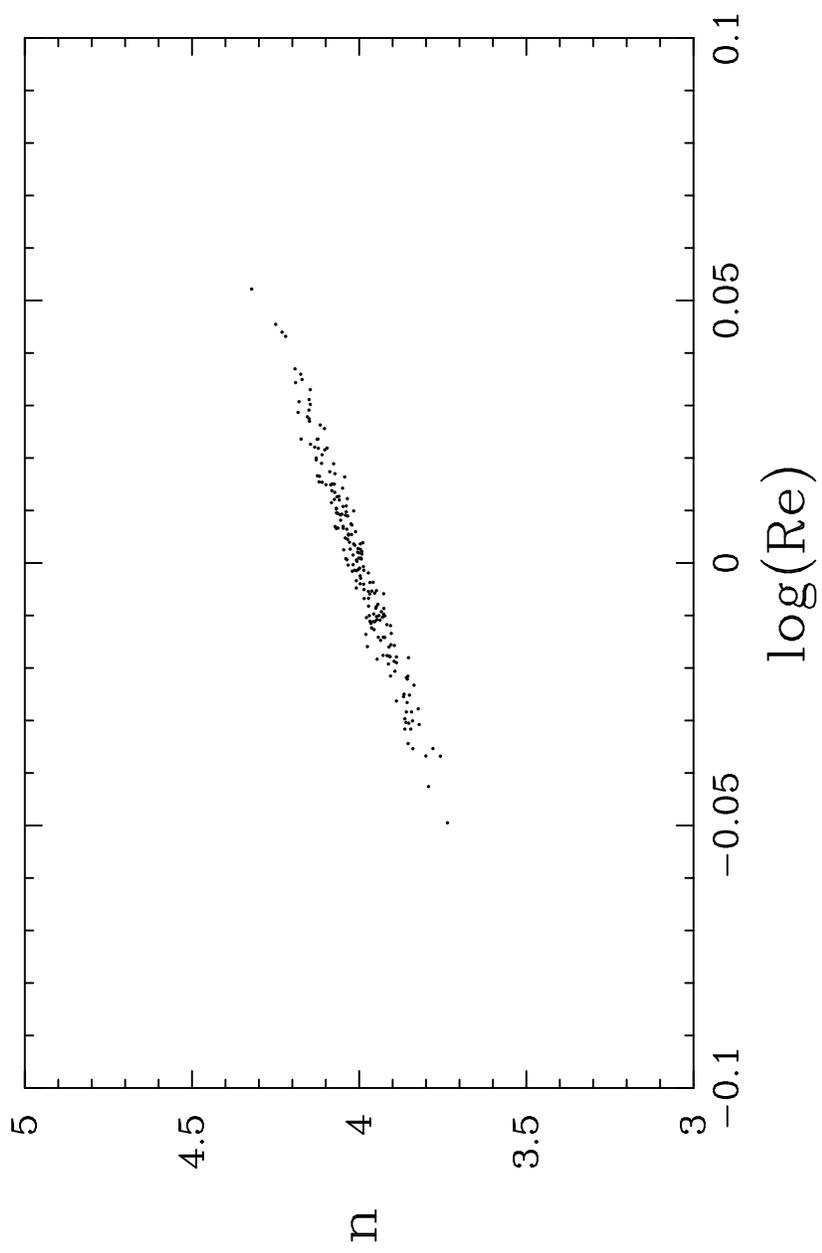
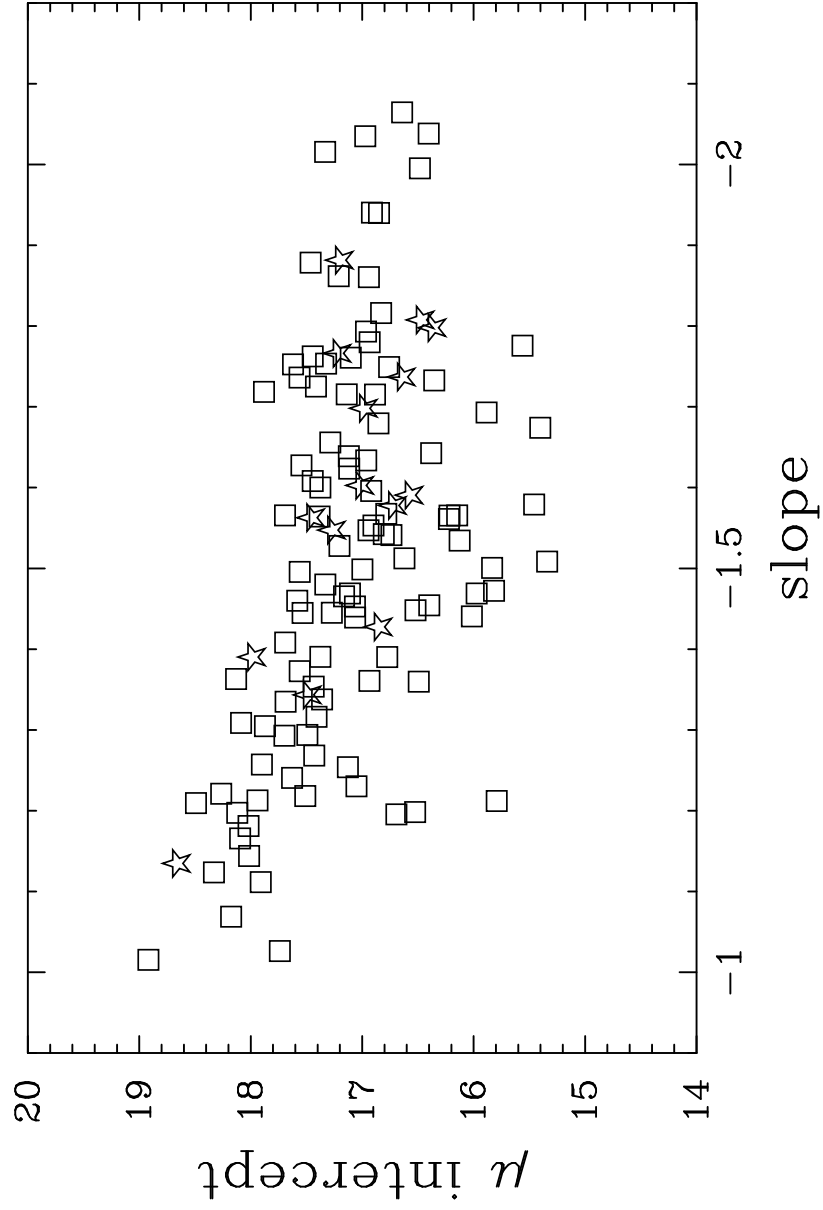
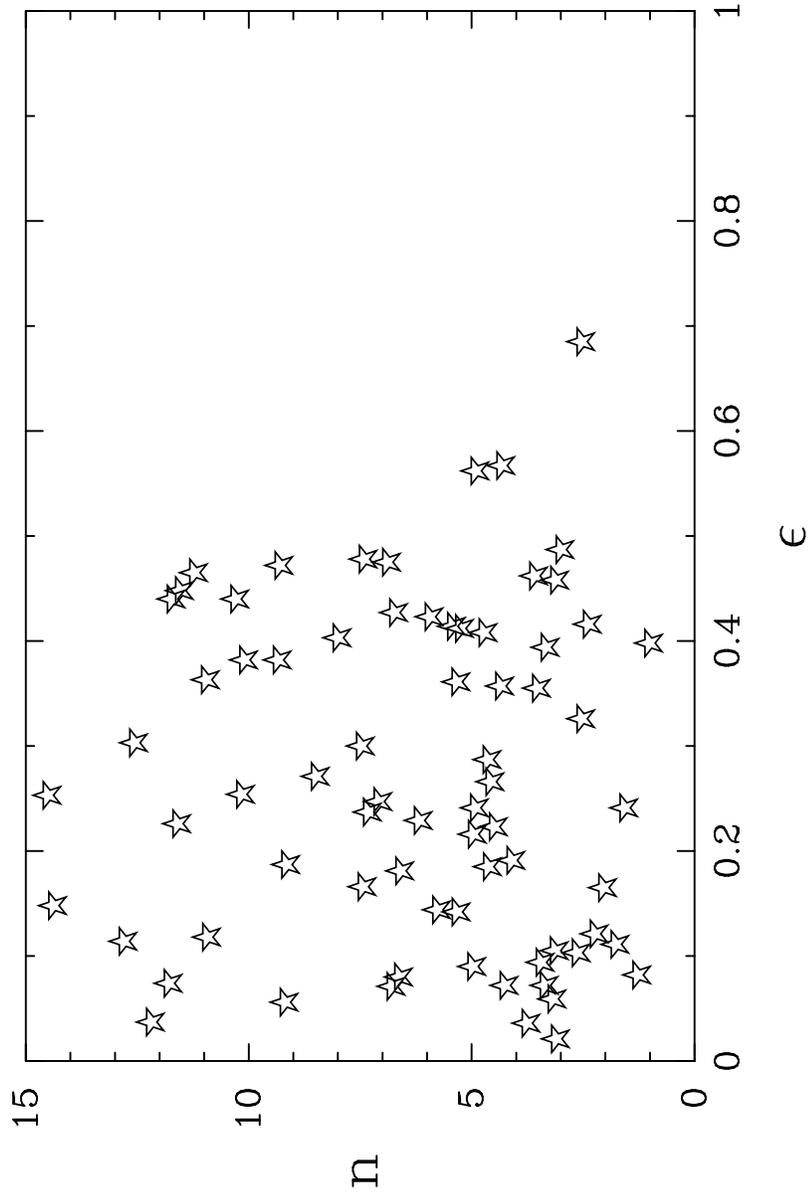
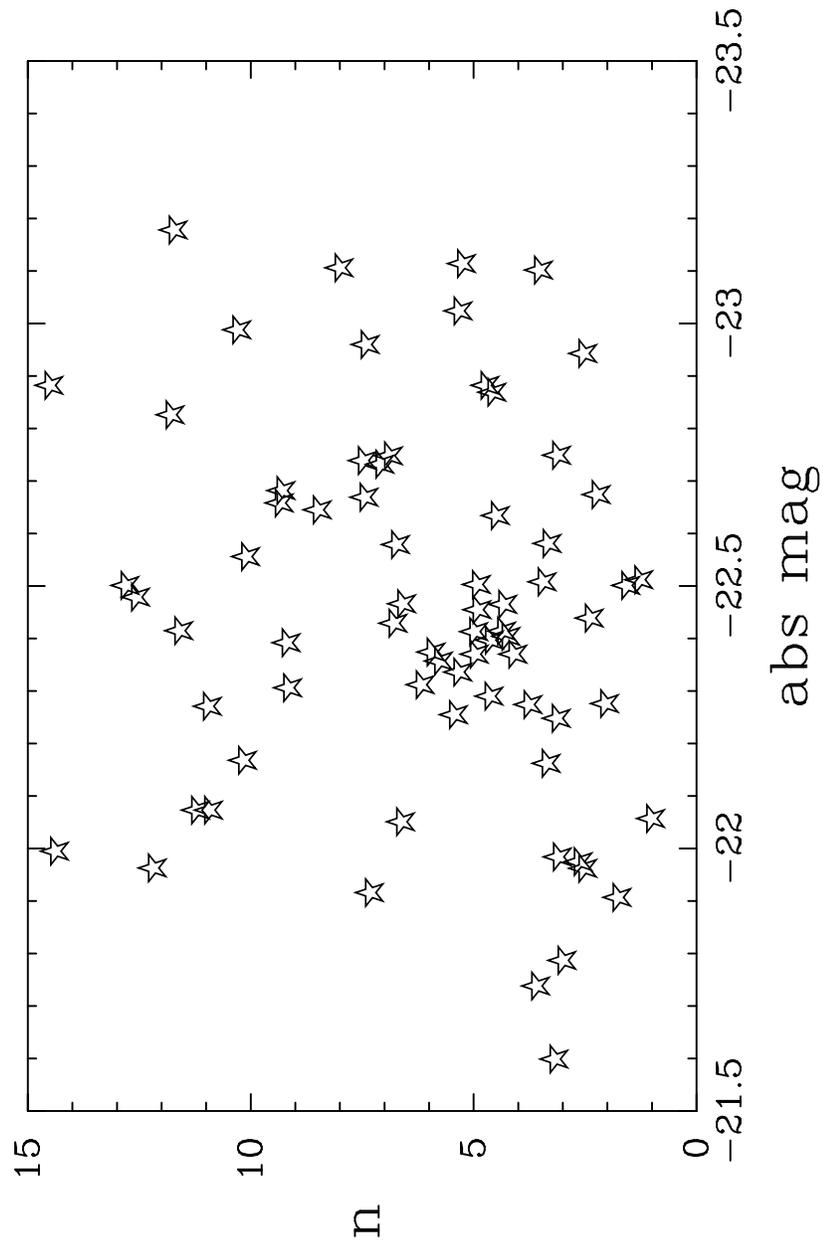
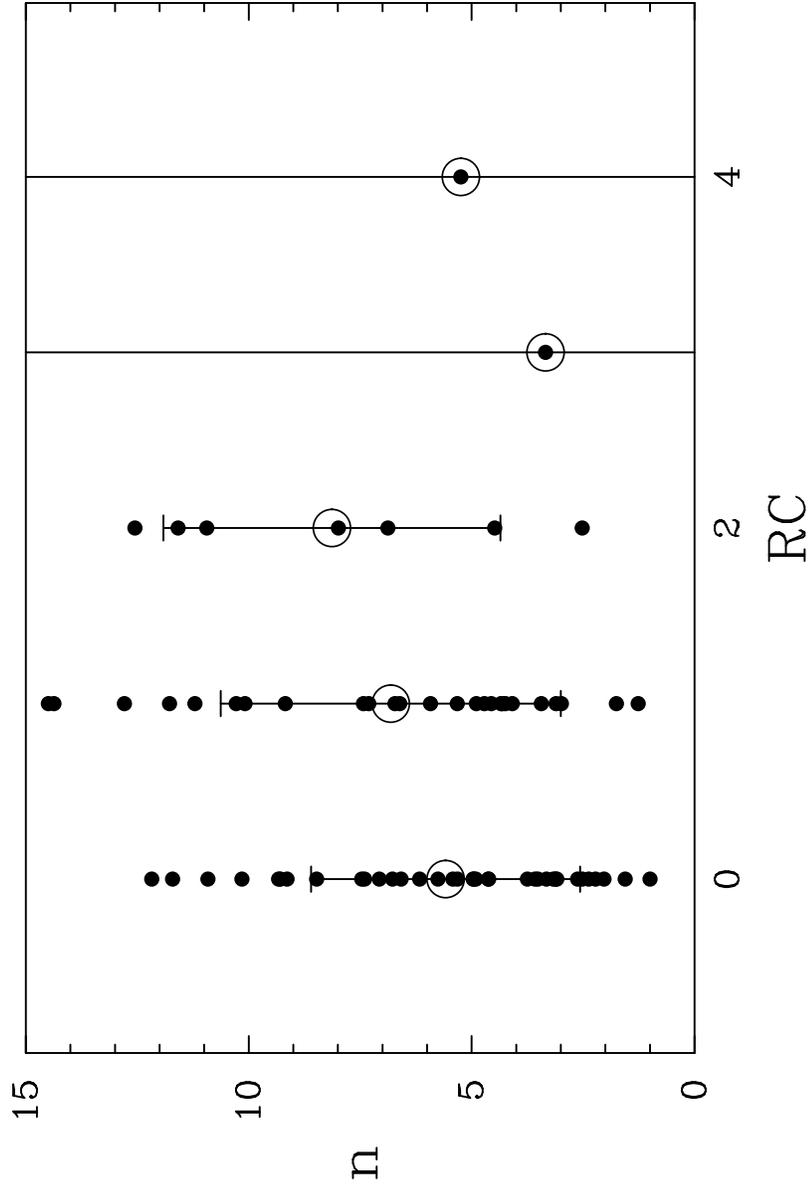
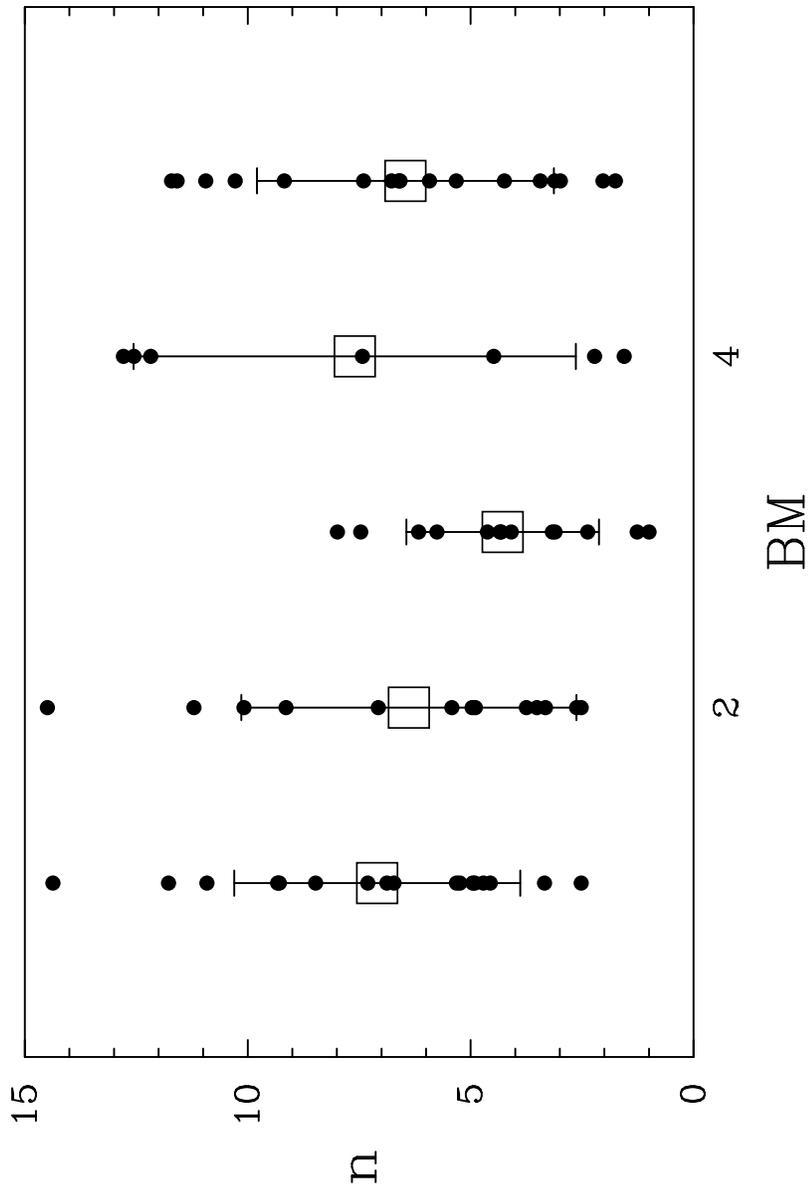
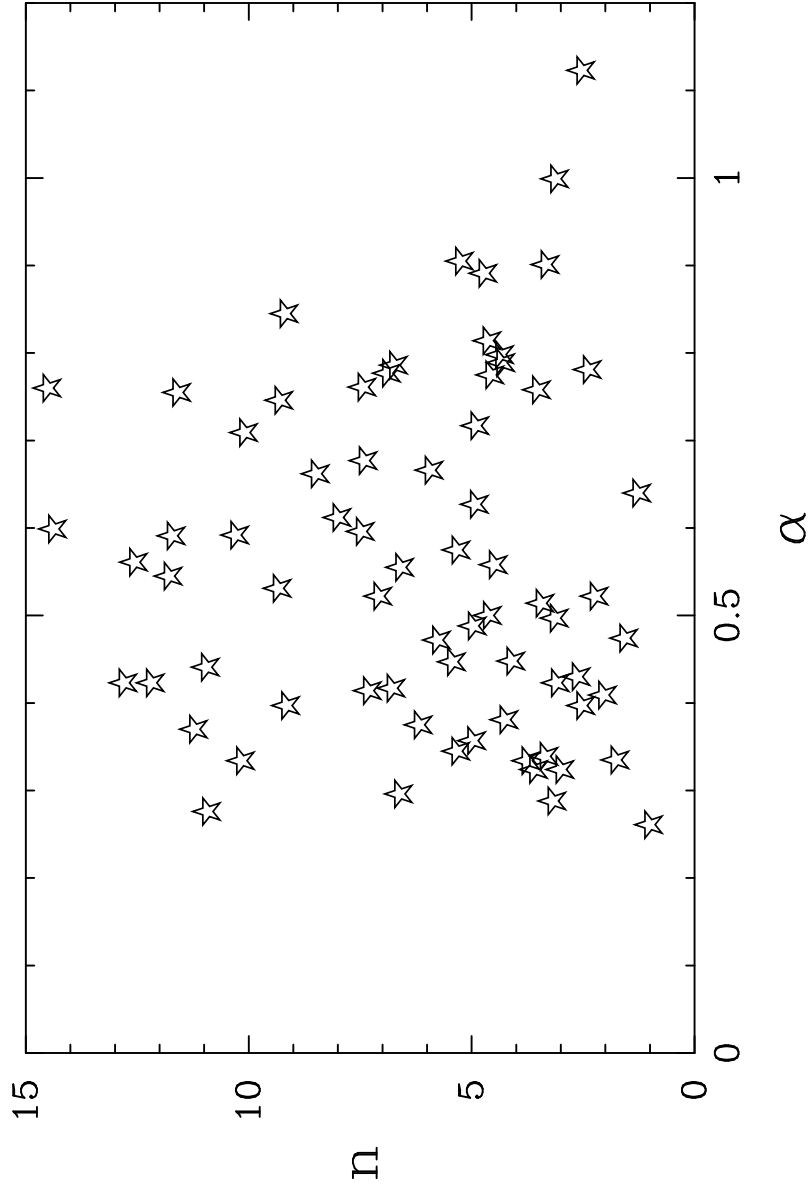
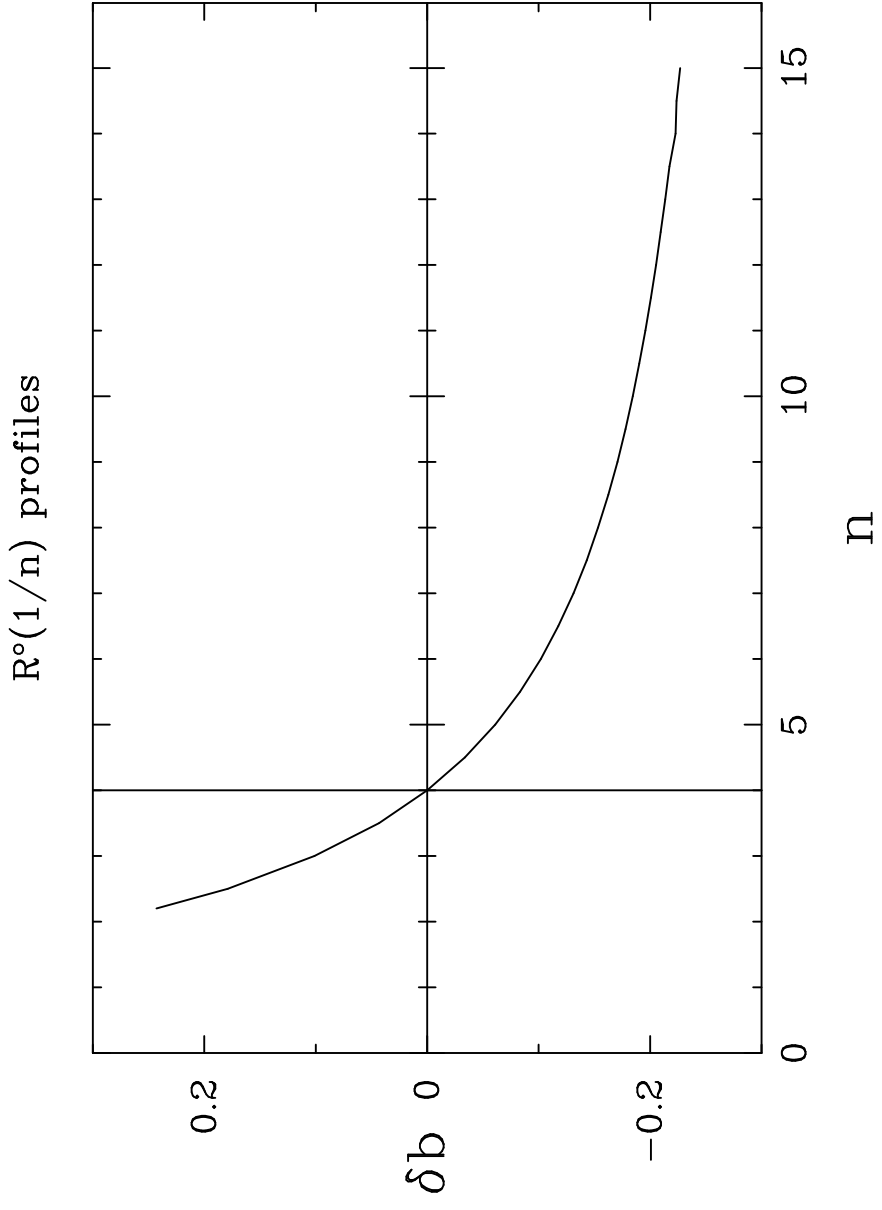
\end{document}